\documentclass[twocolumn,prb,showpacs,preprintnumbers,amsmath,amssymb]{revtex4}% PRB

\usepackage{graphicx}% Include figure files
\pdfoutput=1
\begin{document}

%\preprint{APS/123-QED}

\title{Polarization memory in single Quantum Dots}
\author{E. Poem}
\email{poem@tx.technion.ac.il}
\author{S. Khatsevich}
\author{Y. Benny}
\author{I. Marderfeld}
\author{D. Gershoni}
\affiliation{Department of physics, The Technion - Israel institute of technology, Haifa, 32000, Israel}
\author{A. Badolato}
\author{P. M. Petroff}
\affiliation{Materials Department, University of California Santa Barbara, CA, 93106, USA}

\date{\today}

\begin{abstract}
We measured the polarization memory of excitonic and biexcitonic
optical transitions from single quantum dots at either positive,
negative or neutral charge states. Positive, negative and no
circular or linear polarization memory was observed for various
spectral lines, under the same quasi-resonant excitation below the
wetting layer band-gap. We developed a model which explains both
qualitatively and quantitatively the experimentally measured
polarization spectrum for all these optical transitions.
%Our model uses the full configuration-interaction method, including the
%electron-hole exchange interaction, for calculating the quantum
%dot's confined many-carrier states and the selection rules for
%optical transitions between these states.
We consider quite generally the loss of spin orientation of the
photogenerated electron-hole pair during their relaxation towards
the many-carrier ground states. Our analysis unambiguously
demonstrates that while electrons maintain their initial spin
polarization to a large degree, holes completely dephase.
\end{abstract}


\maketitle

\section{Introduction}
Charge-carriers in semiconductor quantum dots (QDs) are
three-dimensionally confined and quite isolated from their immediate
environment. Therefore, their spin states are relatively protected,
resulting in long lifetimes and slow dephasing rates~\cite{finley}.
As such, they are considered by many as candidates for stationary,
solid-state qubits~\cite{loss,dg,imamoglu}, the building blocks for
quantum information processing~\cite{qip}.

The spin states of charge carriers in semiconductors can be
addressed externally by means of optical
orientation~\cite{zakharchenya}. This possibility establishes, in
principle, external avenues for 'reading', 'writing' and
manipulating these in-matter, stationary
qubits~\cite{ebbens,atature,xu,gerardot,ramsay,press}. Many recent
efforts have been therefore devoted to study the optical properties
of semiconductor quantum dots in general~\cite{dekel,regelman}, and
their polarization sensitive spectroscopy in
particular~\cite{gammon,ware,akopian,ediger,poem}. Correlations
between the polarization of the light which excites QDs
resonantly~\cite{ramsay,paillard} or
quasi-resonantly~\cite{finley,ebbens} and the polarization of the
photoluminescence (PL) that they consequently emit have been studied
both in single~\cite{ebbens,ware,young,favero,bracker} and in
ensembles of QDs~\cite{paillard,borri,tartakovskii,braun,laurent}.
In particular, effects of positive and
negative~\cite{cortez,laurent,kalevich,shabaev} circular and
linear~\cite{gammon,paillard,favero} polarization memory have been
experimentally observed and theoretically discussed
~\cite{cortez,kalevich}. Most of these studies, however, presented
one particular experimental observation, pertaining to a given
charge state, or particular excitation conditions. Thus, the gained
understanding have not been either compared with, or applied to a
wider range of observations.

In this work, we describe comprehensive experimental and theoretical
study of the degree of circular and linear polarization memory (DCPM
and DLPM, respectively) in quasi-resonantly excited single QDs. We
were able to identify and investigate excitonic and biexcitonic
transitions from seven different positive, negative and neutral
charge states of the same QD. The experimentally observed quite rich
polarization memory spectra reveal positively charged spectral lines
with positive DCPM, negatively charged lines with either positive or
negative DCPM and some lines which have no polarization memory at
all. In, addition, we find that none of the spectral lines at this,
quasi-resonant excitation conditions, closely below the wetting
layer bandgap energy, exhibit DLPM.

Our experimental observations are analyzed using a many-carrier,
full configuration interaction (FCI) model~\cite{poem}. We use the
model, which takes into account also the electron-hole exchange
interaction, for calculating the confined many carriers collective
states and optical transitions between them~\cite{poem}.\\
The polarization memory effect is introduced into the model by
allowing only the quasi resonantly excited spin polarized electron
hole pair to lose its spin orientation during its relaxation to the
ground many carrier states. The reasoning behind this assumption is
the vast body of experimental and theoretical evidences that QD
confined ground state charge carriers do not lose their spin
orientation within a typical radiative time scale (~1
nanosecond)~\cite{finley,young,heiss,golovach}.\\
The relaxation to the ground state is followed by radiative
recombination which we straightforwardly calculate by our FCI
model~\cite{poem}.

Comparison between the experimental observations and the theoretical
model yields quantitative agreement with all the obseved spectral
lines. This agreement unambiguously demonstrate that while electrons
memorize their initial spin polarization during their
thermalization, holes completely dephase.
%%%%%%%%%%%%%%%
\section{\label{sec:exp}Experimental methods}
%%%%%%%%%%%%%%%
\subsection{Sample}The studied sample was grown by molecular beam epitaxy on a (001)
oriented GaAs substrate. One layer of strain-induced InGaAs QDs was
deposited in the center of a 285 nm thick intrinsic GaAs layer. The
GaAs layer was placed between two distributed Bragg reflecting
mirrors (DBRs), made of 25 (bottom DBR) and 10 (top DBR) periods of
pairs of AlAs/GaAs quarter wavelength thick layers. This constitutes
a one optical wavelength in matter microcavity for light emitted due
to recombination of QD confined e-h pairs in their respective lowest
energy states.

In order to apply electric fields on the QDs and thereby change
their charge state, a \mbox{p-i-n} structure was formed by n- (p-)
doping the bottom (top) DBR, while leaving the GaAs spacer
intrinsic. In addition, a 10 nm thick AlAs barrier was grown inside
the GaAs spacer between the p-type region and the QDs. This barrier
prolongs the hole's tunneling time into (out of) the QDs at forward
(reverse) bias, with respect to the tunneling time of the electron.
In this way the QDs could have been charged negatively or positively
upon forward or reverse bias, respectively.
%%%%%%%%%%%%%%%
\subsection{Optical characterization}
For the optical measurements the sample was mounted on the cold
finger of a He-flow cryostat, maintaining temperature of about 20K.
A X60 in-situ microscope objective was used in order to both focus
the exciting beam on the sample surface and collect the emitted
light. The collected light was dispersed by a 1~meter monochromator
and detected by an electrically-cooled CCD array detector with
spectral resolution of about 10 $\mu eV$ per one CCD camera pixel.
The polarization of the exciting beam was defined and that of the
emitted light was analyzed by using two sets of two computer
controlled liquid crystal variable retarders and a linear polarizer.
\begin{figure}[tbh]
  \includegraphics[height=0.4\textheight]{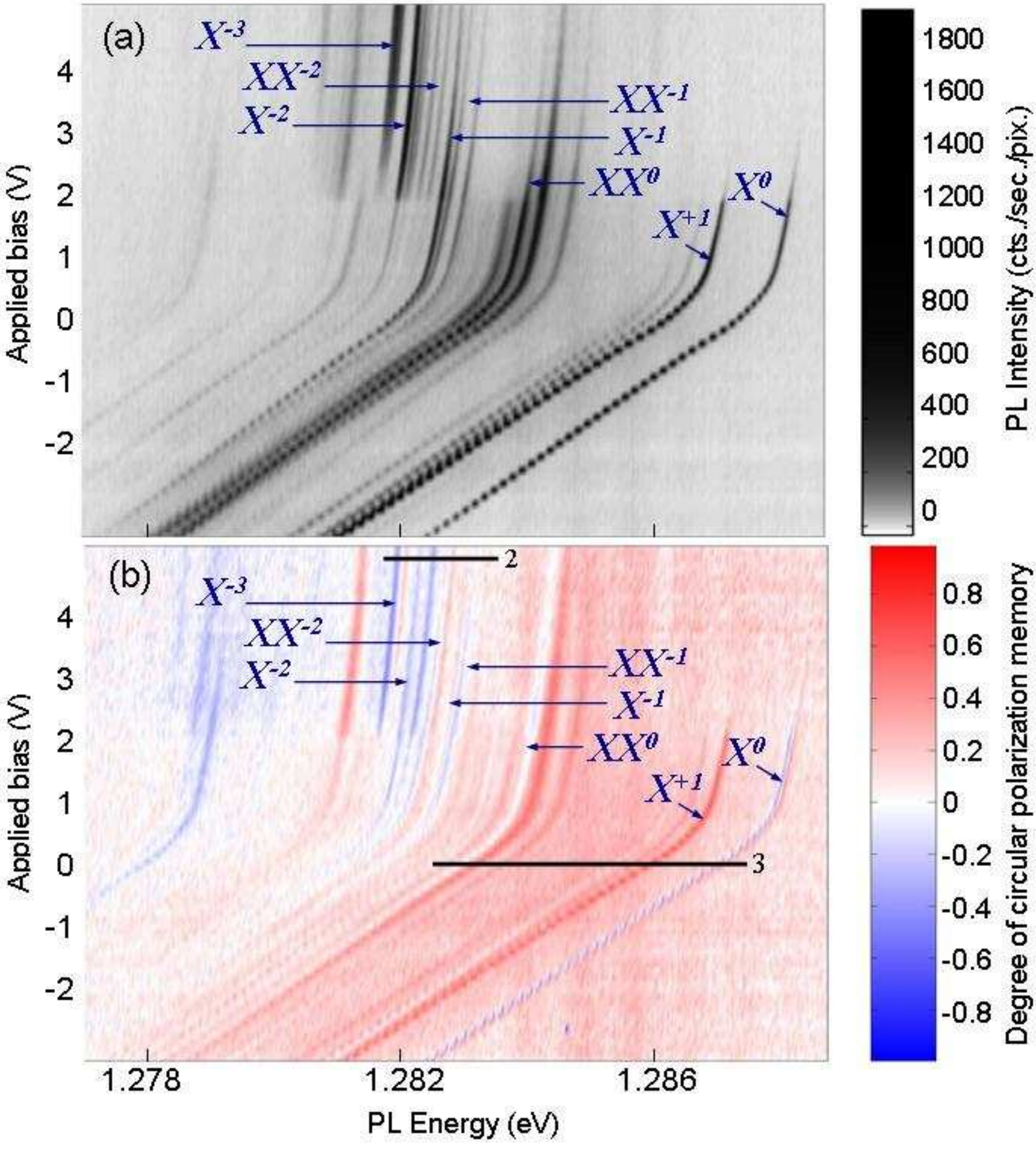}
  \caption{(Color online) Bias dependent PL spectra (a) and
   DCPM (b) from a single QD excited at 1.369 eV. The black horizontal
   lines marked 2 and 3 indicate the bias and spectral ranges from which  Figs. 2 and 3 were obtained.
   }
\label{fig:1}
\end{figure}

In \mbox{Fig. \ref{fig:1}(a)} we present bias dependent
photoluminescence (PL) spectra from one single QD, optically excited
at 1.369 eV. At this energy, a few meV below the bandgap of the InAs
wetting layer, the QDs are quasi-resonantly excited~\cite{ware}. At
reverse bias the spectral lines are red-shifted due to the applied
electric field, and lines due to optical transitions in the presence
of positive charges are enhanced. At forward bias, flat-band
conditions are reached and spectral lines due to the presence of
negative charges appear. The various spectral lines are identified
by their bias dependence, their order of appearance, and by their
polarized fine structures~\cite{poem}. In \mbox{Fig. \ref{fig:1}(b)}
we present the DCPM spectra as a function of the bias. The DCPM is
defined as \mbox{$P_{circ}=(I^+_+-I^+_-)/(I^+_++I^+_-)$}, where
\textit{I} stands for the PL intensity, and the superscript
(subscript) $+$ ($-$) stands for right- (left-) hand circular
polarization of the exciting (emitted) light. Clearly, the DCPM of
each and every spectral line is almost bias independent. While for
all positive lines the DCPM is positive, different negative lines
have different DCPM signs.
\begin{figure}[tbh]
  \includegraphics[height=0.4\textheight]{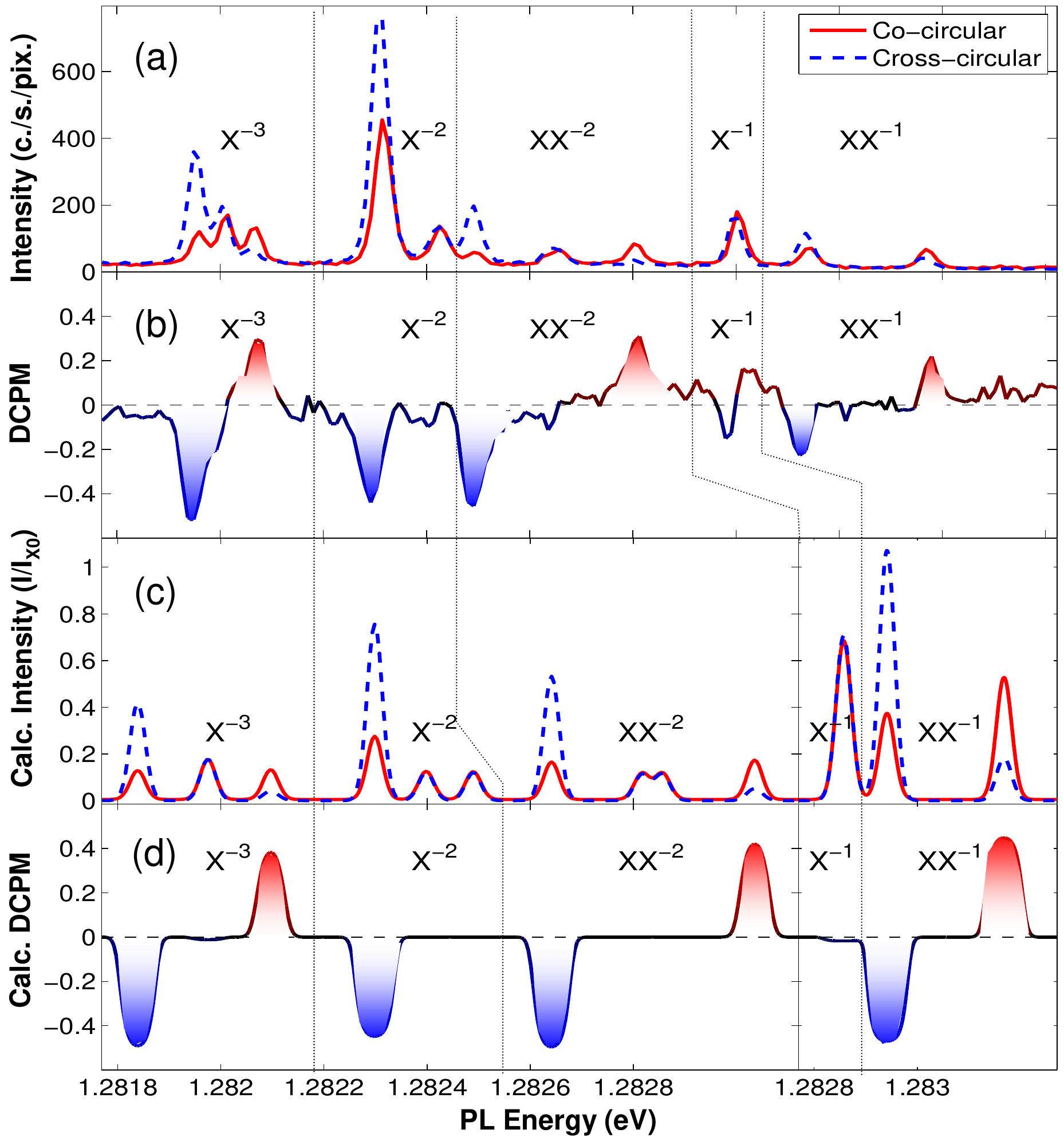}
  \caption{(Color online) (a) measured and (c) calculated polarization sensitive spectra at 4.9 V.
  The solid red (dashed blue) line represents spectrum obtained with co- \mbox{(cross-)} circularly polarized
  excitation and detection: \mbox{$I_{co}=I^+_+$} \mbox{($I_{cross}=I^+_-$)}.
  (b) measured and (d) calculated degree of circular polarization memory.
  The dotted vertical lines are guides to the eye.}
\label{fig:2}
\end{figure}

In \mbox{Fig. \ref{fig:2}(a)} we present spectra obtained at a
forward bias of 4.9 V. At this voltage the QD is negatively charged
with 1 - 3 electrons. The solid red (dashed blue) line represents
the spectrum obtained when the excitation and collection are co-
(cross-) circularly polarized. In \mbox{Fig. \ref{fig:2}(b)} we
present the corresponding DCPM. In \mbox{Fig. \ref{fig:2}} one
clearly observes again that the DCPM sign depends on the specific
optical transition. Some spectral lines show positive memory, like
all the lines associated with positive charge do. Some show no
polarization memory, and some show negative polarization memory.
\begin{figure}[tbh]
  \includegraphics[height=0.4\textheight]{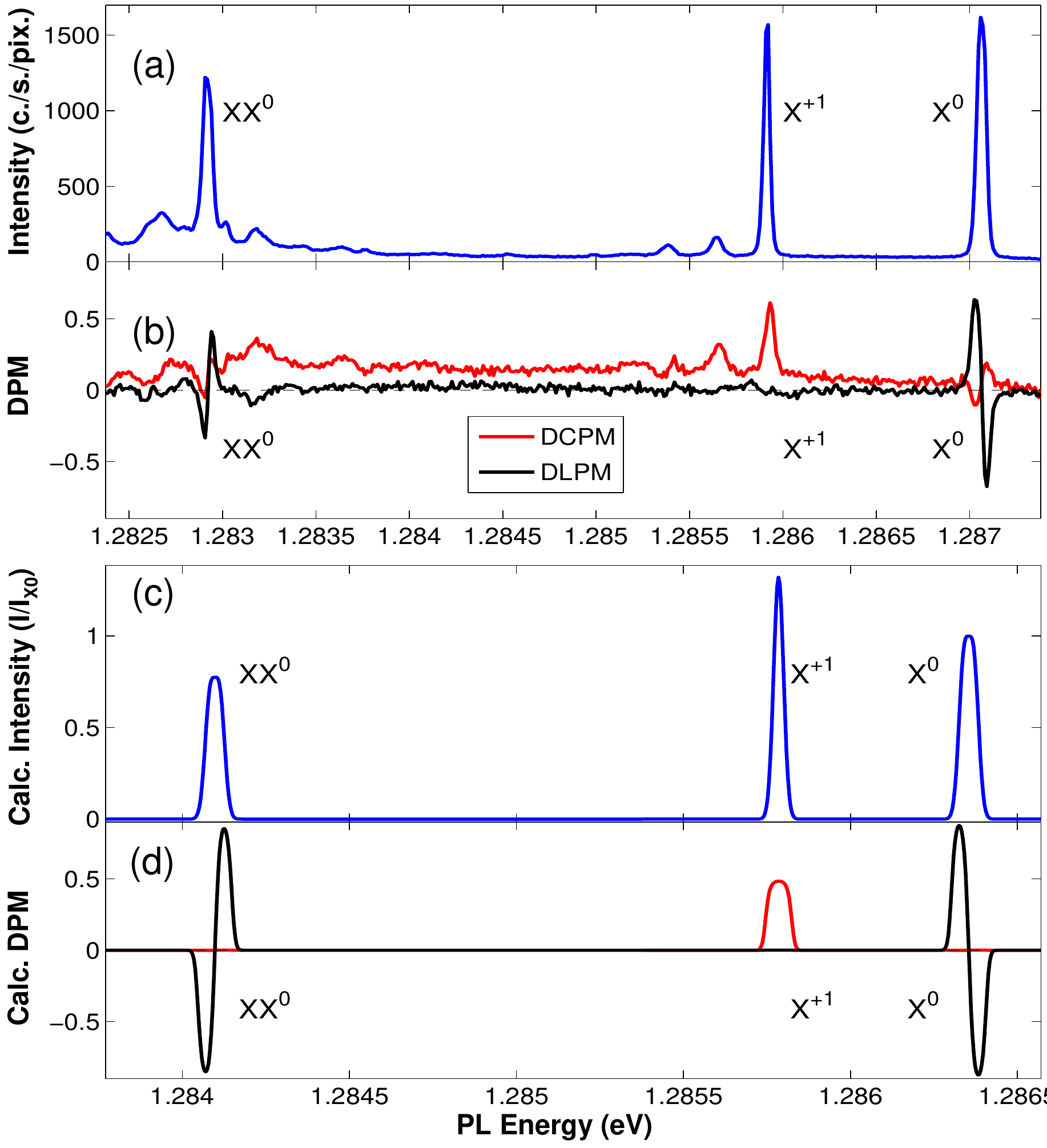}
  \caption{(Color online) (a) measured and (c) calculated unpolarized PL spectra at 0 V.
  (b) measured and (d) calculated degrees of circular (red line) and linear (black line) polarization
  memory.}
\label{fig:3}
\end{figure}

In \mbox{Fig. \ref{fig:3}(a)} we present the spectrum obtained at 0
V. In \mbox{Fig. \ref{fig:3}(c)} we present the measured DCPM and
DLPM. The DLPM is defined as
\mbox{$P_{lin}=(I^H_H-I^H_V)/(I^H_H+I^H_V)$}, where the horizontal
(H) [vertical (V)] direction is determined by the polarization
direction of the lower (higher) energy fine-structure component of
the neutral exciton (X$^0$) line. The X$^0$ line shows no DCPM, and
in total no DLPM either, since its H and V polarized fine-structure
components are equally visible upon H linearly polarized excitation.
We note that the X$^{+1}$ (positively charged exciton) shows strong
positive DCPM but no DLPM.
%%%%%%%%%%%%%%%
\section{\label{sec:theo}Theoretical Model}
In order to explain these observations and to gain further insight
into the phenomenon of polarization memory in optically excited
single semiconductor quantum dots we developed a single-band, full
configuration-interaction model, which includes the electron-hole
exchange interaction (EHEI)~\cite{poem}. We use the model to
calculate the quantum dot's confined many-carrier states and the
selection rules for optical transitions between these states. Prior
to the optical excitation the states within 1 meV from the ground
state of a given number of $N_h$ holes and $N_e$ electrons were
considered to be populated with equal probability. This assumption
is compatible with thermal distribution at the ambient temperature
of the experiment.
%EILON PLEASE CHECK!
We consider the polarized quasi-resonant excitation at a given
polarization by adding an additional electron-hole pair to these
states. The spin state of the additional carriers are defined by
their initial spin polarization, $S_{exc}$, dictated by the
polarization of the exciting light, and by their spin dephasing
during thermalization.

Quite generally, we describe the spin orientation loss by 4
probabilities which apply to each carrier independently. The
probabilities $p_j^{e(h)}$ are for either spin orientation
preservation, $j=0$, or for spin rotations by $\pi$ radians about
the spatial directions x, y, and z for $j=1,2$ and $3$,
respectively. The spin states of the thermalized pair can now be
represented by a 4x4 density matrix in the Hilbert space of the
pair's spin states $
 \uparrow\Uparrow,\uparrow\Downarrow,\downarrow\Uparrow,\downarrow\Downarrow$:
\begin{equation}\label{eq:spin_decay}
  \rho^{th}=\sum_{j,j'=0}^{3}p^e_jp^h_{j'}\sigma^e_j\otimes\sigma^h_{j'}|S_{exc}\rangle\langle S_{exc}|\sigma^{e\dag}_j\otimes\sigma^{h\dag}_{j'}
\end{equation}
where $\sigma^{e(h)}_j$ are the Pauli matrices acting on the
sub-space of electron ($\uparrow$) (hole ($\Uparrow$)) spin states
and $\sigma_0$ is the unit matrix.

If one further assumes that the spin orientation loss (or dephasing)
for both carrier types is isotropic, the number of independent
probabilities to be considered is reduced to two. Thus:
$p_1^{e(h)}=p_2^{e(h)}=p_3^{e(h)}=p^{e(h)}$ and
$p_0^{e(h)}=1-3p^{e(h)}$. We note here that defining these
probabilities in the more frequently used terms of $T_1$ and $T_2$
times~\cite{finley} is straightforward, if the thermalization times
are known.

The additional pair increases the number of charge carriers to
$N_h+1$ holes and $N_e+1$ electrons. The new many carrier states are
restricted to these many carrier states which accommodate the
photogenerated carriers with their spin orientation. For an initial
state $|A\rangle$ of $N_e$ electrons and $N_h$ holes, the resulting
density matrix which defines the states with the additional
thermalized pair is given by

\begin{equation}\label{eq:after_excitation}
  \rho_A=\sum_{\alpha,\beta}
{\rho^{th}_{\alpha\beta}\hat{x}^{\dagger}_{\alpha}|A\rangle\langle
A|\hat{x}_{\beta}}
\end{equation}
where $\hat{x}^{\dagger}_{\alpha}$ is the creation operator of an
electron-hole pair with spin $\alpha$ in any combination single
elctron and single hole spatial states:
\begin{equation}\label{eq:pair_creation}
  \hat{x}^{\dagger}_{\alpha}=\sum_{m,n}
{\hat{a}^{\dagger}_{m,\alpha_e}\hat{b}^{\dagger}_{n,\alpha_h}}
\end{equation}
where $\hat{a}^{\dagger}_{m,\alpha_e}$
($\hat{b}^{\dagger}_{n,\alpha_h}$) is the creation operator of an
electron (a hole) in the single electron (hole) spatial state $m$
($n$), and the spin state $\alpha_e$ ($\alpha_h$), where the spin
state of the electron-hole pair is
$\alpha\equiv\{\alpha_e,\alpha_h\}$.
%$\alpha\in\{\uparrow\Uparrow,\uparrow\Downarrow,\downarrow\Uparrow,\downarrow\Downarrow\}$.

With this description of the $N_e+1$, $N_h+1$ state, we proceed by
projecting it on all energy `ground' states $|G\rangle$ within 1~meV
of the lowest energy level of this number of charge carriers, which
are the states which contribute to photoluminescence. We then
conclude by calculating the energies $\varepsilon$ and intensities
for polarized optical transitions $I_{S_{em}}^G(\varepsilon)$ with
polarization $S_{em}$ from the ground state $|G\rangle$ to states of
$N_h$ holes and $N_e$ electrons~\cite{poem,dekel:ssc}. The $S_{em}$
polarized spectrum for $S_{exc}$ polarized quasi-resonant excitation
is then obtained by summing over all the thermally populated initial
states $|A\rangle$ and over all optically excited $|G\rangle$ states
contributing to the photoluminescence:
\begin{equation}\label{eq:pol_pol_intensity}
    I_{S_{em}}^{S_{exc}}(\varepsilon)=\sum_{G,A}{Tr(\rho_A|G\rangle\langle G|)\cdot I_{S_{em}}^G(\varepsilon)}
\end{equation}
where $\rho_A$ is obtained from $|S_{exc}\rangle$ by
Eq.~\ref{eq:spin_decay} and Eq.~\ref{eq:after_excitation}.

The two probabilities $p^e$ and $p^h$ of Eq.~\ref{eq:spin_decay} can
now be found by comparing the measured DCPM and DLPM to the
calculated ones. The values \mbox{$p^e=1/8$} and \mbox{$p^h=1/4$}
describe very well the observations for this particular
quasi-resonant excitation. These values mean that while the hole
totally loses its polarization during the thermalization, the
electron's degree of polarization is reduced to half. Kalevich
\textit{et al}~\cite{kalevich} previously assumed a similar
situation to successfully explain their observation of negative
circular polarization memory in an ensemble of doubly-negatively
charged QDs.

The calculated spectra for co- and cross- circularly polarized
emission from a negatively charged quantum dot with 1 up to 3
charges were added together to form the calculated polarization
sensitive spectra in Fig. \ref{fig:2}(c).  Both single exciton and
biexciton emissions were included. Gaussian broadening of \mbox{35
$\mu$eV} was assigned for each allowed optical transition. The
obtained calculated DCPM spectrum is presented in \mbox{Fig.
\ref{fig:2}(d)}. By comparing the measured and calculated
polarization sensitive spectra and DCPM, one clearly notes that all
the features of the measured DCPM are given by this simple model. %We
%note that the positive DCPM exhibited by positively charged lines
%(not shown) is well in accordance with this model.
In \mbox{Fig. \ref{fig:3}(c)} we present the calculated spectrum for
the neutral exciton (X$^0$), the neutral biexciton (XX$^0$), and the
singly positively charged exciton (X$^{+1}$). In \mbox{Fig.
\ref{fig:3}(d)} we present the corresponding calculated DCPM (red)
and DLPM (black). The H (V) directions are along the long (short)
semi-axes of the model QD~\cite{poem}. The positive DCPM of the
X$^{+1}$ spectral line and the lack of DCPM from the neutral
excitonic transitions are clearly reproduced by our model. In
addition the model clearly reproduces the experimentally measured
lack of DLPM from all the observed spectral lines at this quasi
resonant excitation energy. We note here, however, that DLPM is
observed in some cases of resonant
excitations~\cite{gammon,paillard,favero}. In these cases, (to be
presented and discussed elsewhere), both carrier types do not
completely lose their initial spin polarization orientation during
the thermalization prior to the recombination.\\

\begin{figure}[tbh]
  \includegraphics[height=0.35\textheight]{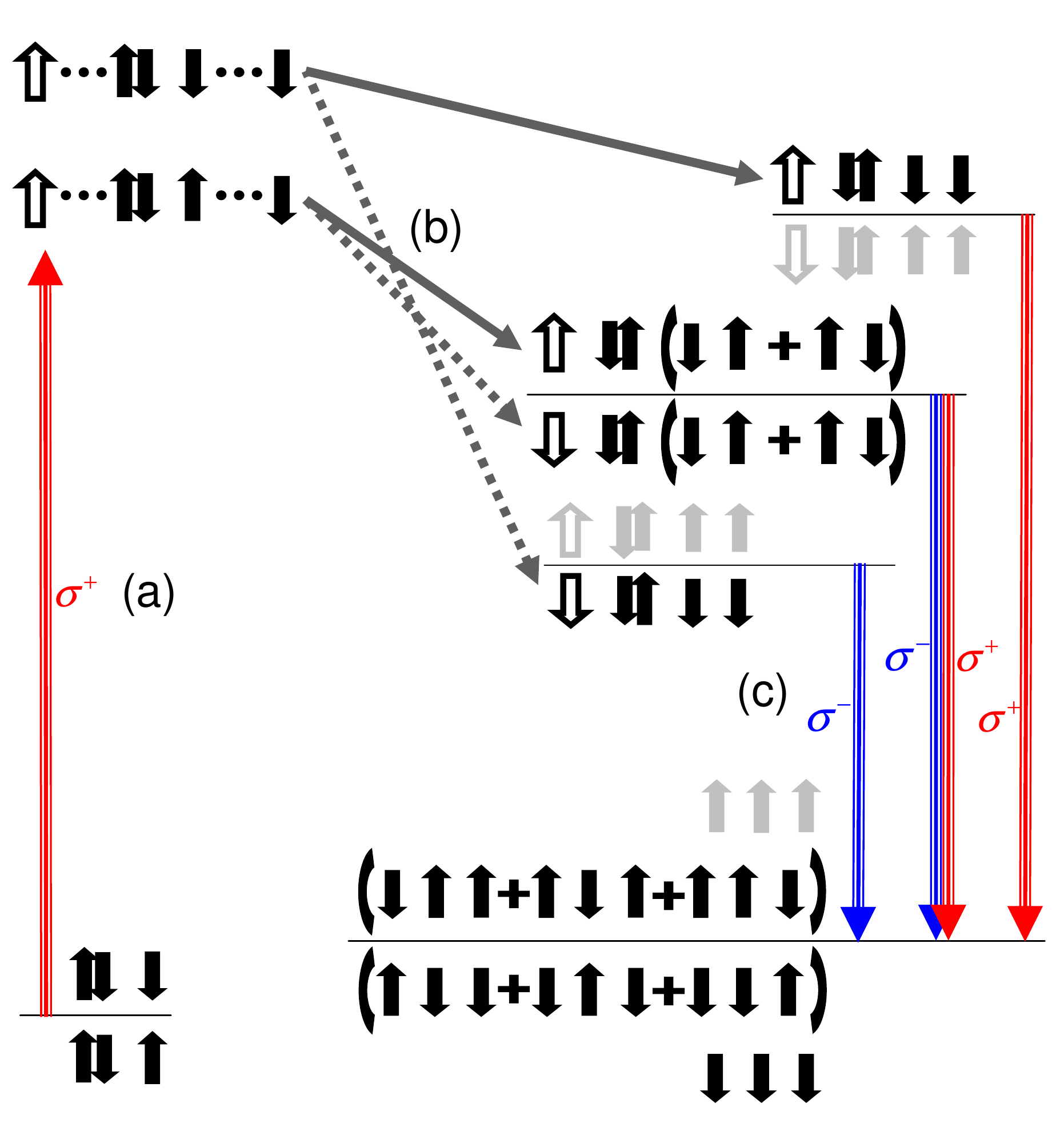}
  \caption{(Color online) Schematic description of the processes which lead to the
  observed DCPM among the X$^{-3}$ spectral lines.
  The symbol $\Uparrow$ (\boldmath{$\downarrow$})
  represents a spin-up (down) hole (electron). the symbols
  are ordered from left to right in increasing energy order ($s$, $p_x$, $p_y$).
  Gray color represent states which do not participate in the
  described process. (a) An electron-hole pair is photogenerated by a quasi-resonant
  $\sigma^+$ polarized excitation and added to three QD electrons residing in their
  ground states. (b) During the thermalization, the hole spin projection along the growth direction
  is either preserved (solid dark-gray arrows), or flipped (dotted dark-gray arrows).
  The lowest (highest) energy levels of the ground $N_e=4$, $N_h=1$  states, is
  reached only if the hole flips (preserves) its spin orientation.
  The intermediate level is reached in both cases.
  (c) All three levels return via radiative recombination of an $s$ shell electron hole pair to
  the same four-fold degenerate $N_e=3$, $N_h=0$ level, giving rise to spectral lines with
  positive, negative and no DCPM, respectively.}
\label{fig:4}
\end{figure}
%
%of the two cross linearly polarized fine-structure components of the
%$^0$ and of those of the XX$^0$ are clearly reproduced by the model.
We identify the main cause of the observed DCPM phenomena as the
isotropic-EHEI induced energetic separation between states where the
electron and hole spins are parallel, and those where they are
anti-parallel. Since circularly polarized excitation always involve
electron-hole pairs with anti-parallel spins, states with (anti-)
parallel spins can be reached only in cases where one (none) of the
carriers flips its spin, yielding negative (positive) circular
polarization memory.
%%% !!!! added new paragraph !!!! %%%
We note that in the case of the doubly negatively charged exciton
(X$^{-2}$), the appearance of negative DCPM for the lower-in-energy
doublet indicates that the energy splitting between the two
components of this doublet is smaller than the radiative width of
the lines~\cite{kalevich,dzhioev}. Consequently, we set this
particular EHEI energy to zero in our model~\cite{poem}.\\
%Zero DCPM occurs either when both the initial and final many-carrier
%states do not include an unpaired electron and an unpaired hole.
%Clearly, in these cases, there is no EHEI. (\textit{e.g.} in the
%case of X$^{-1}$), or where the anisotropic-EHEI causes the dipole
%moment for the optical transition to be linear (\textit{e.g.} in the
%higher-energy doublet of the X$^{-2}$).
As an illustration of the processes involved in the polarization
memory, we schematically describe in \mbox{Fig. \ref{fig:4}} the
case of quasi resonant excitation of the X$^{-3}$ spectral line.
%We turn now to a more qualitative and intuitive discussion of the
%mechanism which leads to either positive, negative or no circular
%polarization memory. The various many carriers states between which
%optical transitions take place can be either degenerate or doublets,
%depending on the relative spin orientations of the electrons and
%holes. Therefore, the radiative transitions can be divided into
%three types. Transitions in which the initial states are split and
%the final states are degenerate, transitions in which the initial
%states are degenerate and the final states are split, and
%transitions in which both initial and final states are degenerate.
%In transitions of the first (second) group, the relation between the
%spin of the hole and the spin of the electrons in the initial
%(final) state: either, parallel, anti-parallel or parallel and
%antiparallel, will determine whether the polarization memory will be
%negative, positive or none, respectively. In transitions of the
%third type, the polarization memory will always be zero. The
%statements regarding the first and third type transitions, are valid
%only if electrons maintain their spin while holes do not. For the
%statement regarding the second type of transitions to be true the
%assumption of spin preserving electronic relaxation is sufficient.
%%%%%%%%%%%%%%%
\section{\label{sec:sum}Summary}
In summary, we measured positive, zero and negative degree of
circular polarization memory in optical transitions from various
negatively charged states of single quantum dots at quasi-resonant
optical excitation. At the same conditions, transition originated
from oddly positively charged states show only positive degree of
circular polarization. All the observed spectral lines do not show
appreciable degree of linear polarization memory. We developed a
model which provides means for calculating polarization memory for
any polarization state of the exciting light and any many carrier
state of a single quantum dot. By applying the model to the case
under study we provide quantitative agreement with all the
experimental observations. The agreement is achieved by two fitting
parameters: the isotropic spin flip probabilities of the
photogenerated electron and hole during their thermalization. We
show that under the conditions of our quasi-resonant excitation,
photogenerated electrons partially preserve their initial spin
orientations, while holes completely dephase.
%%%%%%%%%%%%%%%
\begin{acknowledgments}
The support of the US-Israel binational science foundation (BSF),
the Israeli science foundation (ISF), the ministry of science and
technology (MOST) and that of the Technion's RBNI are gratefully
acknowledged.
\end{acknowledgments}
%%%%%%%%%%%%%%%

%%%%%%%%%%%%%%%

\begin{thebibliography}{00}

\bibitem {finley}
D. Heiss, M. Kroutvar, J. J. Finley, and G. Abstreiter,
%"Progress towards single spin optoelectronics using quantum dot nanostructures",
Solid~State~Commun. {\bf 135}, 591 (2005).

\bibitem {loss}
D. Loss and D. P. DiVincenzo,
%"Quantum computation with quantum dots",
Phys.~Rev.~A. {\bf 57}, 120 (1998).

\bibitem{dg}
D. Gershoni,
%"Long live the spin"
Nature Materials {\bf 5}, 255, (2006).

\bibitem {imamoglu}
A. Imamoglu, D. D. Awschalom, G. Burkard, D. P. DiVincenzo, D. Loss,
M. Sherwin, and A. Small,
%"Quantum Information Processing Using Quantum Dot Spins and Cavity QED",
Phys.~Rev.~Lett. {\bf 83}, 4204, (1999).

\bibitem {qip}
C. H. Bennet and G. Brassard, in \emph{IEEE International Conference
on Computers, Systems and Signal Processing} (IEEE, New York, 1984).

\bibitem {zakharchenya}
F. Meier and B. P. Zakharchenya, Eds., \emph{Optical Orientation}
(North-Holland, Amsterdam, 1984).

\bibitem {ebbens}
A. Ebbens, D. N. Krizhanovskii, A. I. Tartakovskii, F. Pulizzi, T.
Wright, A. V. Savelyev, M. S. Skolnick, and M. Hopkinson,
%"Optical orientation and control of spin memory in individual InGaAs quantum dots"
Phys.~Rev.~B. {\bf 72}, 073307 (2005).

\bibitem {atature}
M. Atat\"{u}re, J. Dreiser, A. Badolato, A. H\"{o}gele, K. Karrai,
and A. Imamoglu,
%
Science {\bf 312}, 551 (2006).

\bibitem {xu}
X. Xu, Y. Wu, B. Sun, Q. Huang, J. Cheng, D. G. Steel, A. S.
Bracker, D. Gammon, C. Emary, and L. J. Sham,
%
Phys. Rev. Lett. {\bf 99}, 097401 (2007).

\bibitem {gerardot}
B. D. Gerardot, D. Brunner, P. A. Dalgarno, P. \"{O}hberg, S. Seidl,
M. Kroner, K. Karrai, N. G. Stoltz, P. M. Petroff, and R. J.
Warburton,
%
Nature {\bf 451}, 441 (2008).

\bibitem {ramsay}
A. J. Ramsay,1 S. J. Boyle, R. S. Kolodka, J. B. B. Oliveira, J.
Skiba-Szymanska, H. Y. Liu, M. Hopkinson, A. M. Fox, and M. S.
Skolnick,
%
Phys. Rev. Lett. {\bf 100}, 197401 (2008).

\bibitem {press}
D. Press, T. D. Ladd, B. Zhang, and Y. Yamamoto,
%
Nature {\bf 456}, 218 (2008).

\bibitem{dekel}
E. Dekel, D. Gershoni, E. Ehrenfreund, D. Spektor, J. M. Garcia, and
P. M. Petroff,
%"Multiexciton Spectroscopy of a Single Self-Assembled Quantum Dot"
Phys. Rev. Lett. {\bf 80}, 4991 (1998).

\bibitem{regelman}
D. V. Regelman, U. Mizrahi, D. Gershoni, E. Ehrenfreund, W. V.
Schoenfeld, and P. M. Petroff,
%"Semiconductor Quantum Dot: A Quantum Light Source
% of Multicolor Photons with Tunable Statistics"
Phys. Rev. Lett. {\bf 87}, 257401 (2001).

\bibitem {gammon}
D. Gammon, E. S. Snow, B. V. Shanabrook, D. S. Katzer, and D. Park,
%
Phys. Rev. Lett. {\bf 76}, 3005 (1996).

\bibitem{ware}
M. E. Ware, E. A. Stinaff, D. Gammon, M. F. Doty, A. S. Bracker, D.
Gershoni, V. L. Korenev, \c{S}. C. B\u{a}descu, Y. Lyanda-Geller,
and T. L. Reinecke,
%"Polarized Fine Structure in the
% Photoluminescence Excitation Spectrum of a Negatively Charged
% Quantum Dot"
Phys. Rev. Lett. {\bf 95}, 177403 (2005).

\bibitem{akopian}
N. Akopian, N. H. Lindner, E. Poem, Y. Berlatzky, J. Avron, D.
Gershoni, B. D. Gerardot, and P. M. Petroff,
%
Phys. Rev. Lett. {\bf 96}, 130501 (2006).

\bibitem {ediger}
M. Ediger, G. Bester, B. D. Gerardot, A. Badolato, P. M. Petroff, K.
Karrai, A. Zunger, and R. J. Warburton,
%
Phys. Rev. Lett. {\bf 98}, 036808 (2007).

\bibitem{poem}
E.~Poem J.~Shemesh, I.~Marderfeld, D.~Galushko, N.~Akopian,
D.~Gershoni, B.~D.~Gerardot, A.~Badolato,  and P.~M.~Petroff,
Phys.~Rev.~B {\bf 76}, 235304 (2007).

\bibitem {paillard}
M. Paillard, X. Marie, P. Renucci, T. Amand, A. Jbeli, and J.-M.
G\'{e}rard,
%"Spin relaxation quenching in semiconductor quantum dots"
Phys.~Rev.~Lett. {\bf 86}, 1634 (2001).

\bibitem {young}
R. J. Young, S. J. Dewhurst , R. M. Stevenson, P. Atkinson, A. J.
Bennett, M. B. Ward, K. Cooper, D. A. Ritchie, and A. J. Shields,
%"Single electron-spin memory with a semiconductor quantum dot"
New~J.~Phys. {\bf 9}, 365 (2007).

\bibitem {favero}
I. Favero, G. Cassabois, C. Voisin, C. Delalande, Ph. Roussignol, R.
Ferreira, C. Couteau, J. P. Poizat, and J.-M. G\'{e}rard,
%"Fast exciton spin relaxation in single quantum dots"
Phys.~Rev.~B. {\bf 71}, 233304 (2005).



\bibitem {bracker}
A. S. Bracker, E. A. Stinaff, D. Gammon, M. E. Ware, J. G. Tischler,
A. Shabaev, Al. L. Efros, D. Park, D. Gershoni, V. L. Korenev, and
I. A. Merkulov,
%"Optical pumping of the electronic and nuclear spin of single
% charged tunable quantum dot"
Phys.~Rev.~Lett. {\bf 94}, 047402 (2005).

\bibitem {borri}
P. Borri, W. Langbein, S. Schneider, U. Woggon, R. L. Sellin, D.
Ouyang, and D. Bimberg,
%
Phys.~Rev.~Lett. {\bf 87}, 157401 (2001).

\bibitem {tartakovskii}
A. I. Tartakovskii, J. Cahill, M. N. Makhonin, D. M. Whittaker,
J.-P. R. Wells, A. M. Fox, D. J. Mowbray, M. S. Skolnick, K. M.
Groom, M. J. Steer, and M. Hopkinson,
%"Dynamics of Coherent and Incoherent Spin Polarizations
% in Ensembles of Quantum Dots"
Phys.~Rev.~Lett. {\bf 93}, 057401 (2004).

\bibitem {braun}
P.-F. Braun, B. Eble, L. Lombez, B. Urbaszek, X. Marie, T. Amand, P.
Renucci, O. Krebs, A. Lema\^{\i}tre, P. Voisin, V. K. Kalevich, and
K. V. Kavokin,
%"Spin relaxation of positive trions in InAs/GaAs quantum dots:
% the role of hyperfine interaction"
Phys.~Stat.~Sol. (b) {\bf 243}, 3917 (2006).

\bibitem {laurent}
S. Laurent, M. Senes, O. Krebs, V. K. Kalevich, B. Urbaszek, X.
Marie, T. Amand, and P. Voisin,
%"Negative circular polarization as a general property of n-doped self
% assembled InAs/GaAs quantum dots under nonresonant optical excitation"
Phys.~Rev.~B. {\bf 73}, 235302 (2006).

\bibitem {cortez}
S. Cortez, O. Krebs, S. Laurent, M. Senes, X. Marie, P. Voisin, R.
Ferreira, G. Bastard, J.-M. G\'{e}rard, and T. Amand,
%"Optically driven spin memory in n-doped InAs-GaAs quantum dots"
Phys.~Rev.~Lett. {\bf 89}, 207401 (2002).

\bibitem{kalevich}
V.~K.~Kalevich, I.~A.~Merkulov, A.~Yu.~Shiryaev, K.~V.~Kavokin,
M.~Ikezawa, T.~Okuno, P.~N.~Brunkov, A.~E.~Zhukov, V.~M.~Ustinov and
Y.~Masumoto, Phys.~Rev.~B {\bf 72}, 045325 (2005).

\bibitem {shabaev}
A. Shabaev, E. A. Stinaff, A. S. Bracker, D. Gammon, Al. L. Efros V.
L. Korenev, and I. Merkulov,
%"Optical pumping and negative luminescence polarization
% in charged GaAs quantum dots"
Phys.~Rev.~B. {\bf 79}, 035322 (2009).

\bibitem{heiss}
D. Heiss, S. Schaeck, H. Huebl, M. Bichler, G. Abstreiter, J. J.
Finley, D. V. Bulaev, and D. Loss,
% \textit{et al},
%
Phys. Rev. B. {\bf 76}, 241306(R) (2007).

\bibitem{golovach}
V. N. Golovach, A. Khaetskii, and D. Loss,
%
Phys. Rev. Lett. {\bf 93}, 016601 (2004).

\bibitem {dekel:ssc}
E. Dekel, D.V. Regelman, D. Gershoni, E. Ehrenfreund, W.V.
Schoenfeld, and P.M. Petroff,
% "Radiative lifetimes of single
%  excitons in semiconductor quantum dots:  Manifestation of the
%  spatial coherence effect "
Solid~State~Commun. {\bf 117}, 395 (2001).

\bibitem {dzhioev}
R. I. Dzhioev, B. P. Zakharchenya, E. L. Ivchenko, V. L. Korenev,
Yu. G. Kusraev, N. N. Ledentsov, V. M. Ustinov, A. E. Zhukov, and A.
F. Tsatsul'nikov,
%"Fine structure of excitonic levels in quantum dots"
JETP Lett. {\bf 65}, 804 (1997).
%Pis'ma~Zh.~Eksp.~Teor.~Fiz. {\bf 65},766

%\bibitem {Khaetskii:PRB00}
%A. V. Khaetskii and Y. V. Nazarov,
%%"Spin relaxation in semiconductor quantum dots"
%Phys.~Rev.~B {\bf 61}, 12639 (2000).
%
%\bibitem {Merkulov:PRB02}
%I. A. Merkulov \textit{et al},
%%"Electron spin relaxation in semiconductor quantum dots"
%Phys.~Rev.~B. {\bf 65}, 205309 (2002).
%
%\bibitem {Woods:PRB02}
%L. M. Woods \textit{et al},
%%"Spin relaxation in quantum dots"
%Phys.~Rev.~B. {\bf 66}, 161318(R) (2002).
%
%\bibitem {Dzhioev:PSolSt98}
%R. I. Dzhioev \textit{et al},
%%"Optical orientation of donor-bound ecxitons in nanosized InP/InGaP nano islands"
%Phys.~Solid~State {\bf 40}, 1587 (1998).
%
%
%\bibitem {Kalevich:PRB01}
%V. K. Kalevich \textit{et al},
%%"Spin redistribution due to Pauli blocking in quantum dots"
%Phys.~Rev.~B {\bf 64}, 045309 (2001).
%
%
%\bibitem {Paillard:PSS00}
%M. Paillard \textit{et al},
%%"Spin repolarization due to Pauli blocking in quantum dots"
%Phys.~Stat.~Sol. (b) {\bf 221}, 71 (2000).
\end{thebibliography}
\end{document}